**Trustworthy Cross-Border Interoperable Identity System for Developing Countries**

Ayei E. Ibor[1], Mark Hooper[2], Carsten Maple[3], Gregory Epiphaniou[4]

[1,2,3] Trustworthy Digital Infrastructure for Identity Systems, The Alan Turing Institute, United Kingdom

[4]WMG, University of Warwick, United Kingdom.

aibor@turing.ac.uk, mhooper@turing.ac.uk, cmaple@turing.ac.uk, gregory.epiphaniou@warwick.ac.uk

**Abstract**

Foundational identity systems (FIDS) have been used to optimise service delivery and inclusive economic growth in developing countries. As developing nations increasingly seek to use FIDS for the identification and authentication of identity (ID) holders, trustworthy interoperability will help to develop a cross-border dimension of e-Government. Despite this potential, there has not been any significant research on the interoperability of FIDS in the African identity ecosystem. There are several challenges to this; on one hand, complex internal political dynamics have resulted in weak institutions, implying that FIDS could be exploited for political gains. On the other hand, the trust in the government by the citizens or identity holders is habitually low, in which case, data security and privacy protection concerns become paramount. In the same sense, some FIDS are technology-locked, thus interoperability is primarily ambiguous. There are also issues of cross-system compatibility, legislation, vendor-locked system design principles and unclear regulatory provisions for data sharing. Fundamentally, interoperability is an essential prerequisite for e-Government services and underpins optimal service delivery in education, social security, and financial services including gender and equality as already demonstrated by the European Union. Furthermore, cohesive data exchange through an interoperable identity system will create an ecosystem of efficient data governance and the integration of cross-border FIDS. Consequently, this research will seek to identify the challenges, opportunities, and requirements for cross-border interoperability in an African context. Our findings show that interoperability in the African identity ecosystem is vital to strengthen the seamless authentication and verification of identity holders for inclusive economic growth and widen the dimensions of e-Government across the continent.

## 1. Introduction

One of the Sustainable Development Goals (SDGs) of the United Nations is to provide legal identity for all citizens, which also includes birth registration by 2030 (World Bank Group, 2018). Primarily, identification allows individuals to exercise their rights to be identified and legally recognised while the government and private sector depend on it for effective service delivery. In this sense, a national identity system is crucial to the identification and authentication of natural or legal persons and provides a means for citizens to access critical services including participation in formal political, social, and economic life (Gelb and Metz, 2018). Achieving sustainable development is a key consideration of most national governments and forms the basis for inclusive and responsible identification schemes at the foundational level.

Fundamentally, national identification schemes are the pivot for social security, immigration, financial and economic inclusion, healthcare, voting, gender equality, transportation, and education (Colbern and Ramakrishnan, 2018). Consequently, identity management plays a key role in promoting e-government by bringing services closer to the people. Modern identification systems are robust tools for the delivery of transparent administration, reduction in fraud and leakages for social benefits, providing adequate security for the citizens, extracting accurate biographic data for effective economic planning, and responding to natural disasters (Atick, 2016). These seeming benefits are yet to solve the identification crisis in most developing nations as about 850 million people across the globe are yet to have access to a valid identity credential, mostly in sub-Saharan Africa and South Asia (The World Bank, 2023a). According to the World Bank Identification for Development (ID4D) dataset (The World Bank, 2023a), most people without official identification are residents of low-income countries. These sets of persons also include marginalised and vulnerable groups consisting of children, whose births were not documented in the civil registry of the affected country, women in rural areas with no access to digital services, and adults below the age of 25 years.

Particularly, digital identification systems can improve how the public and commercial sectors provide services and lay the groundwork for new markets, services, and systems, such as e-government, cashless transactions, and the digital economy. Identification systems must, however, have high levels of coverage and inclusion within the population, be resilient to fraud and error, operate within a governance framework that protects personal data, fosters trust and accountability, and facilitates end-user control to fulfil their potential for facilitating sustainable development and increasing public sector efficiency (African Union, 2020; Bandura and Ramanujam, 2021). There is a growing demand for digital identity to be mutually recognised and portable between countries in the modern digital age through an interoperability framework and in the context of regional and global integration and migration, which can be facilitated through trust and standards.

Interoperability, in this context, refers to the ability of disparate foundational identity systems to exchange data through seamless communication of identification and authentication information. These exchanges must be trustworthy or inherently secure, available, and reliable. Although there are complex internal political dynamics in most developing countries that result in weak institutions, and the lack of trust in government by the citizens including the notion of security and privacy protection concerns, legislation, cross-system compatibility, vendor-locked system design principles and unclear regulatory provisions for data sharing, our findings show that interoperability is vital to widen the dimensions of e-Government. In this paper, the challenges, opportunities, and requirements for trustworthy cross-border interoperability are discussed. The rest of the paper is organised as follows; in Section 2, the background to the study is given while in Section 3, a review of literature is discussed. Section 4 presents the methodology of the research. In Section 5, the discussion of findings is presented, and the conclusion and future work is given in Section 6.

## 2.    Background

## 2.1    Overview of identity management

According to Luong and Park (2023), an identity management system is useful for managing the identity attributes of users. To establish and prove an identity, such identity must first be created through a well-defined registration process. In a foundational identity management system, such a registration process involves the collection, storage, and usage of identity attributes (World Bank Group, 2016). The processes of collecting, storing, and using identity attributes are useful for the verification and validation of ID holders whenever they request services from service providers. Relying parties also depend on these identity attributes to authenticate, authorise, and verify ID holders during transactions or access to a system (Kiourtis et al., 2023). Generally, identification is a key prerequisite for development as it provides the pivot for all categories of transactions and service delivery geared towards inclusive economic growth as summarised in Figure 1.

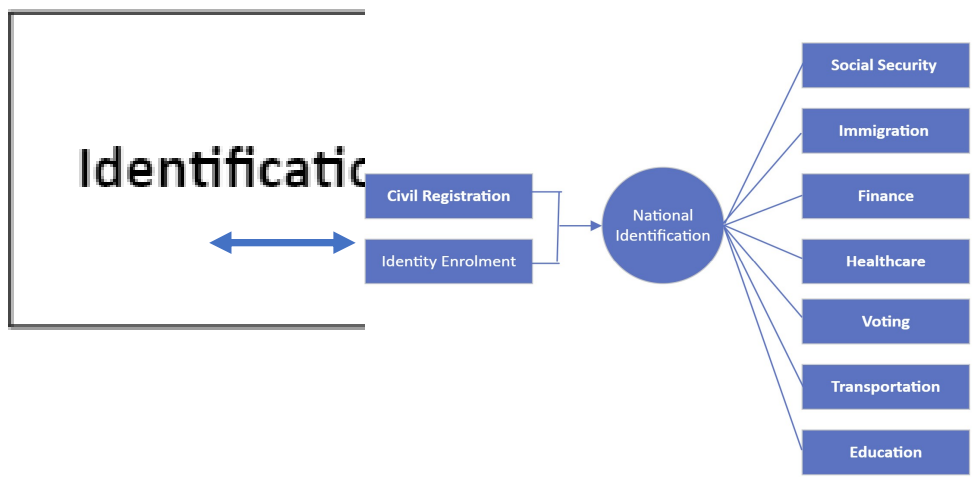

Figure 1: Identification as a key prerequisite for development.

## 2.2 Trustworthiness in digital identity management

Building robust, secure, and reliable identity management systems is a crucial challenge worldwide. Fathalla et al. (2023) argued that trustworthiness in digital identity management ensures that identification data maintains its integrity, security, privacy, and confidentiality. The consequences of identity fraud or theft are discussed in Hummer and Rebovich (2023); Irvin-Erickson (2023), and Walters (2023). With the increasing rate of attacks on digital identities (Pöhn and Hommel, 2023), privacy concerns are becoming more paramount. To this effect, the use of privacy-preserving technologies that do not reveal the real identities of ID holders is being considered (Luong and Park, 2023; Yin et al., 2022; Tang, Ma, and Cheng, 2023).

Typically, a trustworthy system can be relied upon to authenticate and communicate identity attributes without which such information may not be trusted by identity and service providers including relying parties. When an identity system is compromised, it poses a direct threat to the digital identities of ID holders. Significantly, foundational identity systems must be trustworthy since they are part of a nation's critical infrastructure that contributes to inclusive economic growth and e-Government.

## 2.3 Requirements for Interoperability

To achieve the interoperability of foundational identity systems, requirements must be identified at each stage of the identity management process. In this work, we have identified six processes for scoping requirements to achieve a fully functional interoperable identity system in developing countries. These processes include:

i)      Identity creation and management
ii)     Proof of identity
iii)    Creating and issuing a credential
iv)    Issuing a derived credential
v)     Managing the identity credential lifecycle, and
vi)    Granting access to an ID holder.

## 2.4      Challenges and opportunities of interoperability in the African identity ecosystem

The African identity ecosystem is fragmented with vendor-locked systems that are only accessible within the borders of each country (Gelb and Metz, 2018). One of the challenges of interoperability in this context is the trust issues among African nations including their foundational identity systems (Manda and Backhouse, 2016; Domingo and Teevan, 2022). Trust is a key consideration to interoperability as government-to-citizen, government-to-government, government to business interactions are performed over the Internet in an interoperable identity ecosystem.   Similarly, the increase in the use of vendor-neutral technologies for the verification and validation of ID holders at cross-border points is envisaged to introduce more security and privacy concerns.  Moreover, cross-border digital trust requires a secure and reliable environment. Connecting identity systems can create more complex environments for conducting digital transactions and interactions across the continent (African Union, 2020). Challenges in infrastructure, and disparities in social structures, norms, and behaviour, also affect the perception of privacy, security, and trust by stakeholders.

Other challenges include advancements in cutting-edge technology that change how personal data is gathered and analysed from various, unrelated sources including consent management for data sharing, the reluctance of various governments to invest in privacy-enhancing technologies, streamline security policies and legislation, and establish new acts for the protection of the privacy and confidentiality of identification data. Conversely, there are several opportunities for interoperability as it promotes vendor neutrality using common standards in an identity ecosystem. In the same sense, interoperability enforces data integrity by ensuring that each identity system provides a single source of truth for identification data and reduces identity fraud for e-Government services (The World Bank, 2023b; Domingo and Teevan, 2022). With an interoperable continent-wide identity ecosystem, new markets, digital services, and applications are possible, thus enabling innovation and new use cases to widen the dimensions of e-Government.

## 3.      Review of Literature

Although interoperability has the potential to create a seamless identity ecosystem for data exchange, verification, and validation of digital identities, and expand the reach of national governments for effective service delivery, there are very few research-based approaches for cross-border interoperable identity systems. Backhouse and Halperin (2009) posited that the challenge of establishing interoperable systems is enormous, which does not only consider the technical linking of databases and systems. To this effect, they proposed a three-fold conception of interoperability for identity management systems viz-a-viz technical, legal, and regulatory components to enhance data sharing in the provision of e-government.

The factors militating against the full interoperability of federated identity management systems are studied in Catuogno and Galdi (2014). They argued that the tendency for per-site authentication and authorisation of ID holders culminates in huge overhead for both the identity/service providers and the ID holders as each site stores different credentials of the user. Therefore, the authors presented Shibboleth as the de facto standard for identity management and point of access to providers of information (PAPI) as a solution that leverages the joining of federations and translation of protocols during cross-federation authentication and authorization (AA) sessions.

Sharma and Panigrahi (2015) proposed a roadmap useful to plan and implement the capabilities of interoperability in e-government solutions. The roadmap considered the notion of knowledge sharing among key stakeholders based on vital legal, regulatory, technical, and organisational components that can foster the interoperability of e-government services. One significant limitation of this research is its inability to explore cross-country differences in legislation and regulatory requirements, as it focused only on the inputs from stakeholders in India. Also, Kotzé and Alberts (2017) proposed a baseline conceptual model to achieve an e-government interoperability framework. The model considered the technical, legislative, social, and political environments of South Africa to serve as a guideline for enterprises that are evolving towards e-governance.

Similar studies were carried out by Kanagwa et al. (2018), who investigated the relevance of a national enterprise architecture to support several e-government systems including the systems for the registration of persons in Uganda through semantic interoperability that is achieved based on a set of related ontologies. Still, in both studies, country-specific technical, legislative, social, and political considerations were made, which indicate the absence of cross-border considerations to interoperability, Saputro et al. (2020) discussed the Estonian X-Road as an e-governance solution for secure data exchange and the interoperability of information systems across nine countries. This notwithstanding, this research did not categorise whether the countries investigated are developed or developing to ascertain the level of inclusion or exclusion required for facilitating e-governance.

In Hölbl, Kežmah and Kompara (2023), the interoperability and compliance issues in eIDAS are discussed. The authors asserted that eIDAS has limited applicability in the public sector since the regulation does not address new market demands with the added complexity of private online providers connecting to the eIDAS network. Further, the issue of the isolation and inflexibility of notified eID solutions in member states makes it difficult for eIDAS to support a variety of use cases. Domingo and Teevan (2022) discussed the interoperability of cross-border payment solutions to expand trade in Africa. They argued that the success of the African Continental Free Trade Area (AfCFTA) depends largely on the cross-border interoperability of low-value instant payment systems such as mobile money. Consequently, the authors posited that such a policy will enhance inclusion and long-lasting benefits to small-scale businesses.

Additionally, Masiero (2023) argued that the effectiveness of digital identity schemes depends on platform features, which must be considered to comprehend the true extent of harm that digital identification might cause due to interoperability. Furthermore, Benaddi et al. (2023)

identified the challenges in data sharing and the interoperability of e-government systems. Their focus was on the technical interoperability of e-government entities to enhance collaboration in the use of public data. However, their approach does not consider cross-border perspectives to data sharing including the identification of entities that consume these data.

## 4.  Methodology

To underscore the importance of interoperability in developing countries, an investigation and comparison of current interoperability solutions in the identity ecosystem was performed. This investigation identified the current limitations of existing solutions and provided the basis for our findings, which are relevant to achieving interoperability for foundational identity systems in developing nations. As a guide, we have presented an approach for the investigation of requirements in Figure 2.

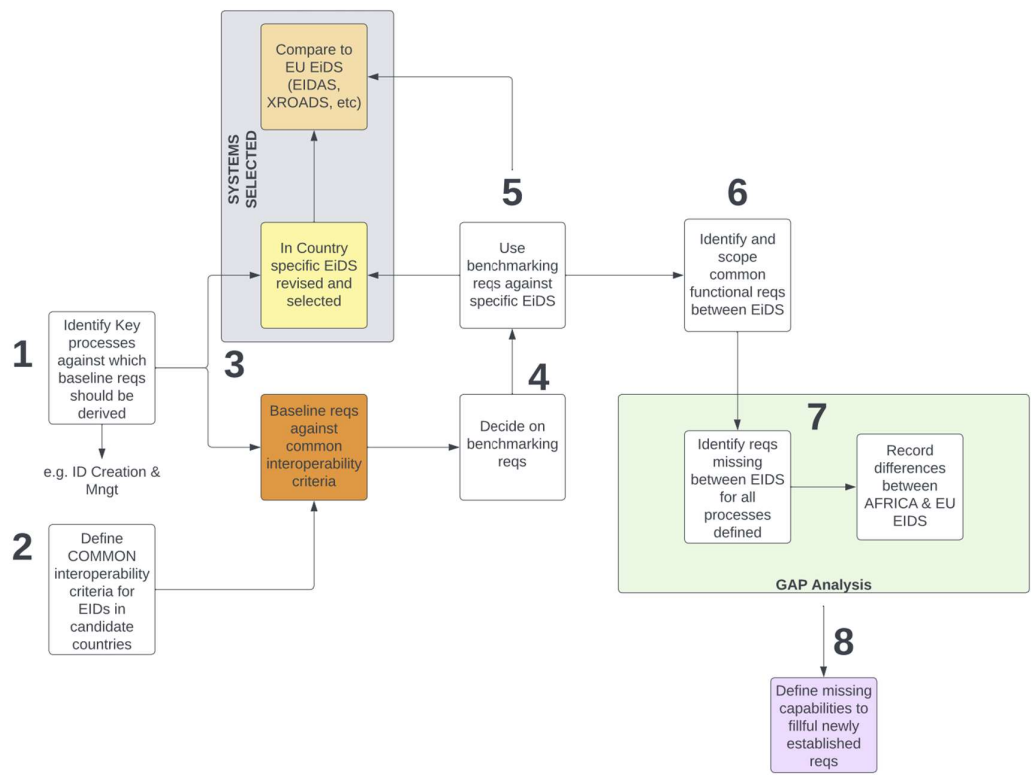

Figure 2: Methodology for scoping requirements for interoperable identity systems.

The approach in Figure 2 is an easy stepwise guide. The steps include:

i)  Identify key processes against which baseline requirements should be derived: these processes are already identified in this work.

ii)  Define common interoperability criteria for eID schemes in candidate countries: this is an important step that must be fulfilled based on the identified key processes of step 1.

iii) Baseline requirements against common interoperability criteria, select and compare country-specific eID schemes with current interoperability solutions such as eIDAS and X-Road: this step allows for the identification of common functionalities among eID schemes relative to defined common interoperability criteria.

iv) Decide on the benchmarking requirements: it is important to decide on the benchmarking requirements to adequately scope the functional requirements useful for a fully functional interoperable system.

v) Use benchmarking requirements against specific eID schemes: using benchmarking requirements against specific eID schemes is useful for identifying missing requirements and functionalities.

vi) Identify and scope common functional requirements between eID schemes: this is a significant step that ensures that all requirements for interoperability are adequately captured.

vii) Perform requirements gap analysis by identifying missing requirements between eID schemes for all processes defined and recording the differences between eID schemes: the missing requirements must be identified in this step concerning the processes defined in step 1. All significant differences between eID schemes must be recorded (e.g. in Africa and Europe).

viii) Define missing capabilities to fulfil newly established requirements: it may be necessary to define missing capabilities in the investigated eID schemes that are relevant to the identified requirements to realise an interoperable identity system. However, this may not be the case where such eID schemes fulfil most of the baseline requirements for interoperability.

## 4.1 Current Interoperability Solutions

For the creation of effective, long-lasting, and usable identity ecosystems, interoperability is essential. Two major interoperability solutions in the identity ecosystem are discussed in this research. These include the Estonian X-Road and the electronic identification and trust services (eIDAS) of the European Union. X-Road is a data exchange layer solution that implements the interoperability of information systems. It allows organisations to exchange data through a secure channel over the Internet. This solution is centrally managed but supports a distributed exchange layer that provides a standardised and secure channel for producing and consuming services (X-Road, 2023a, 2023b).

X-Road enhances the confidentiality, integrity, and interoperability between organisations that rely on it for data exchange (Bakhtina et al., 2022; Saputro et al., 2020). Jointly implemented by Estonia and Finland, it has been adopted by several countries and organisations (Saputro et al., 2020). As stated in Solvak et al. (2019), X-Road establishes online connections between service providers and data registries (such as the Population Register, Health Insurance Register, etc.). The citizens who use the X-Road system only give their information to the government once, and the public authority then stores and exchanges the information among itself via the X-Road system (Bhattarai et al., 2019).

By signing the messages with the X-Road member's signature key and using a mutually authorised Transport Layer Security (TLS) channel, security servers guarantee the integrity and secrecy of the exchanged messages. By recording the exchanged messages and routinely timestamping the message logs, the signed communications' long-term evidential value is protected. To obtain information on the validity of certificates and timestamp-signed messages, the security servers communicate with trust services. In terms of message exchange, the trust service calls are asynchronous (X-Road, 2023c, 2023d).

To exchange messages between a service client and a service provider, three protocols are used. These include X-Road message protocol, X-Road message transport protocol, and OCSP response retrieval protocol (X-Road, 2020). When an interested party such as an organisation joins an X-Road ecosystem, certificates issued by a reputable Certification Authority (CA) are used to verify the identification of each organization and security server. Each security server serves as the technical entry point that manages access control on the organisation level during the data exchange process between registered X-Road members (European Commission, 2023a; X-Road, 2023a). The combination of timestamping and a digital signature ensures that data delivered via X-Road cannot be disputed.

On the other hand, eIDAS is a European Union's framework to ensure that electronic transactions between businesses, citizens, and public agencies are safer and more efficient, regardless of the European country they take place in. It is established by a European Regulation that was implemented in 2014. By introducing a common framework for eID and trust services, the eIDAS regulation makes it easier to supply business services across the EU. It encourages interoperability across the 28 EU nations, making certain that nations mutually recognise each other's electronic identities and trust services across borders (Mocanu et al., 2019).

The goal of electronic identification is to completely transform how customers engage with online services. The Member States of the EU may choose to identify citizens electronically. A small number of Member States have created national programs to provide their citizens with electronic identity (eID), with greatly diverse architectures (Lips, Bharosa and Draheim, 2020). National systems, therefore, vary not only in the volume of citizen data they process but also in the degree of data protection they provide to this data.

As claimed by Cuijpers and Schroers (2014) and Hölbl, Kežmah and Kompara (2023), businesses and customers can more easily access services or conduct commercial transactions by using electronic identity, or eID, to identify who they are (identification process) and demonstrate that they are who they claim to be (authentication process). Similarly, the regulation stipulates that it will be necessary for all EU nations to accept notified eID systems from other nations by September 2018. When conducting electronic transactions, especially those between firms and clients who are based in another EU country, trust services attempt to improve the trust of EU residents and enterprises (Sharif et al., 2022).

The trust services in eIDAS as discussed in ANSSI (2023) and European Commission (2023b) include Electronic Signature (eSignature), Electronic Seal (eSeal), Electronic Timestamp (eTimestamp), Website Authentication Certificates (WACs), and Electronic Registered Delivery Service (eDelivery). eIDAS has been exploited by both the public and private sectors in the EU. Some of the sectors that have benefited from the regulation include the financial services, online retail, transport, and professional services sectors. One of the largest potential beneficiaries of eID and trust services is the financial services industry due to the possibility of enormous commercial opportunities and enhanced cross-border services (Cuijpers and Schroers, 2014; European Commission, 2023b). To meet rising client demand for online services as well as stricter compliance requirements, the identification, authentication, and safeguarding of transactions in the financial services sector are becoming increasingly digitised.

## 4.2    Comparison of X-Road and eIDAS as Interoperability Solutions

The comparison of X-Road and eIDAS is given in Figure 3. This comparison is performed based on identity creation and management, proof of identity, creating and issuing identity credentials, issuing a derived credential, managing the identity credential lifecycle, granting access to an identity holder, data exchange mechanism, security of identification data, and privacy of identification data.

| Process | X-Road | eIDAS |
|---------|--------|-------|
| Identity creation and management | X-Road does not enforce any end-user identity scheme. | Enforces identity schemes through the notification of national eID schemes. |
| Proof of identity | End-user identities are not verified or validated. | Verifies and validates end-user identities such as the eIDs of citizens or ID holders. |
| Creating and issuing identity credentials | Issues credentials to X-Road members and security servers (SS). Does not issue credentials to citizens or ID holders | Creates and issues identity credentials to citizens or ID holders for mutual recognition of eIDs. |
| Issuing a derived credential | Does not issue derived credentials to citizens. | Issues derived credentials to citizens, which are part of the notified eID scheme. |
| Managing the identity credential lifecycle | Manages identities of security servers, organisations and information systems (members) using a certificate authority (CA). The CA issues and revokes certificates to/of members. | Manages identities of member states using a certification body designed by the European Commission to ensure that each member creates qualified electronic signatures and qualified electronic seals. |
| Granting access to an identity holder | Access rights management is based on an authorisation framework using organisation and service level identifiers. | Qualified web authentication certificates (QWAC) are used to authenticate the identification of the natural or legal persons to whom they have been issued, as well as the names of the relevant websites. |
| Data exchange mechanism | Data is exchanged through message routing, which uses organisation and service level identifiers. These identifiers are mapped to the physical network locations of the services. Non-repudiation of data is guaranteed through timestamping and digital signatures. Cross-border data exchange is achieved through the federation of X-Road ecosystems. | Data exchange is achieved through an electronic registered delivery service (eDelivery), which allows the user to send data electronically. eDelivery provides proof of sending and delivery of the data to curb the risk of loss, theft, damage, or unauthorised modifications of the data. The content of the communication between eIDAS nodes is carried out with cryptographically secure SAML messages. |
| Security of identification data | Authentication keys assigned to a Security Server (SS) are used to establish cryptographically secure communications with other SSs and TLS is used to secure messages transmitted over the public Internet. | SAML is used to protect the confidentiality of the person identification data, the authenticity and integrity of the person identification data, and the secure identification of communication endpoints. Transport Layer Security (TLS) is used to protect the communication between the client's browser and server over the Internet e.g., connection via HTTP (HTTPS). |
| Privacy of identification data | Signing keys are assigned to the SS clients and used to sign the exchanged messages. | Certificates for SAML signing and encryption of messages in the eIDAS network are exchanged through signed SAML metadata. |

Figure 3: Comparison of X-Road and eIDAS

**5.     Discussion of Findings**

The limitations of the current interoperability solutions, that is, eIDAS and X-Road show that there is a plethora of issues and complexities, which are yet to be addressed to achieve interoperability at the foundational level of identity management. Country-specific legislation and vendor-locked systems in developing countries add to these complexities.

From the findings, the main limitations of the current interoperability solutions can be summarised as follows:

i.    Each registry implements its interfaces independently based on the use of a proprietary protocol equivalent to the technology in use. This results in the implementation of new interfaces, sometimes from scratch for new or evolving services.

ii.   There are different trust levels based on the implemented security architecture for verifying the integrity and authenticity of the data.

iii.  Incompatible certification authorities arise from different trust levels and legal systems of the trust service providers.

iv.   Interoperability is based on the signing of bilateral agreements, which are non-trivial and can result in several discrepancies in regulations and policies.

v.    There is no provision that the network must be accessible to private entities and as such may build inter-government competition on trust services e.g., eIDAS.

vi.   There is differentiation in notified eID schemes and authentication mechanisms leading to re-identification for public services, healthcare, or financial transactions e.g., eIDAS.

These limitations imply that the interoperability of identity systems requires open standards with strong legal, regulatory, and governance structures. Also, there must be mechanisms to mitigate risks to the security and privacy of identification data including consent considerations for data use or sharing by the ID holder as outlined in Alamillo et al. (2023) and Srinivas, Das and Kumar (2019). From the findings, developing countries must also make provisions for a single, consolidated, and standardised view of civil registrations and identification data that constitute a single source of truth. This will enhance the onboarding of citizens and create a robust verification and validation process that does not require several levels of authentication that may increase the overhead of the identity system.

Additionally, findings showed that the standardisation of the structure and attributes of identity data such as name, date of birth, email, and several other relevant attributes to conform with the W3C recommendation should be a key consideration for interoperability. Enforcing the unicity and singularity of identification data will also ensure that ID holders do not have multiple identities that can hamper interoperability. Similarly, to enable the secure exchange of data or identity assertions, it was also found that developing countries must establish trust relationships through federation protocols that can foster interoperability.

Interoperability portrays tremendous benefits to e-Government. We found that interoperability widens the dimensions of e-Government in cross-border identity management and data services. It also helps to provide open and accessible digital public services including systems and processes that allow people to move freely within the developing countries while also utilising public services outside their country of origin. Interoperability also helps to create sustainability, and economies of scale as demonstrated by X-Road and eIDAS (Hoffmann and Solarte-Vasquez, 2022; McBride et al., 2019; Schmidt and Krimmer, 2022).

We found that while X-Road provides trustworthy data exchange using security servers that allow its members to communicate directly, it does not perform the verification and validation of the identification data that is part of the data exchange. There are also concerns about the limited amount of notified eID schemes under eIDAS, which builds on the limited scope of the eID schemes and the lack of relevant public services. These concerns underpin the need for the review of these interoperability solutions to underscore the notion of cross-border verification and validation of identification data for seamless data exchange. Building on the evidence from various sources as discussed in the Literature, an architecture for trustworthy cross-border interoperability is proposed in this work. This architecture is represented in Figure 4.

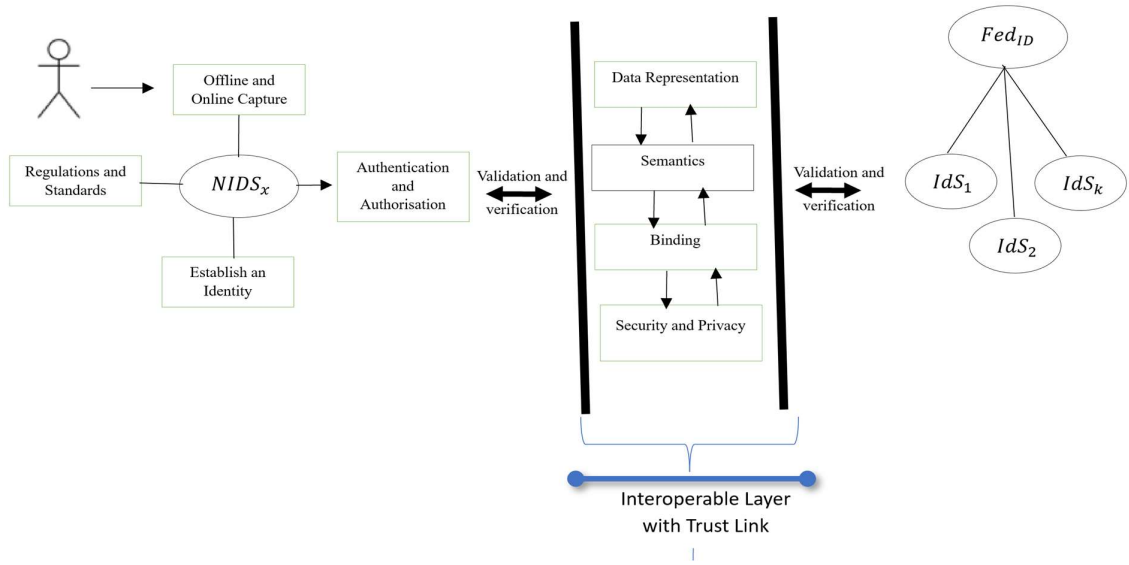

Figure 4: Proposed Architecture for Trustworthy Cross-Border Interoperability

From Figure 4, each citizen or prospective ID holder undergoes identity enrolment, which can include civil registration through online or offline capture procedures. The enrolment of the citizen is based on regulations and standards of the foundational identity system of the issuing country represented in Figure 4 as $NIDS_x$. At the issuance of the ID credential, the citizen or ID holder presents such a document for establishing and proofing his/her identity to a requesting service provider, identity provider or relying party. Authentication and authorisation services are then used to complete the identity proofing process to allow an ID holder access to a service or resource at the point of access.

In a cross-border use case, where a foundational identity system $NIDS_x$ communicates with another system, say, in a federated identity ecosystem $Fed_{ID}$, or where an ID holder $U_{ID}$ requests for a service from the latter, then the processes of verifying and validating the identity of $U_{ID}$ should be an integral component of an interoperable identity system unlike in X-Road where such processes are the functions of the service provider/consumer. Verification and validation ensure that the claimed identity is true and belongs to the claimant at the time of the request and throughout service delivery. To achieve this, there is the need to have a trustworthy link that considers the representation of the data, semantics, binding, and the security and privacy of the identification data of $U_{ID}$. The relevance of the trust link at the interoperable layer is to ensure that the identity data as well as the requested resource or service maintains its integrity, security, privacy, and confidentiality throughout data exchange.

We propose that the representation of the identification data must be data format agnostic as obtainable in X-Road using simple object access protocol (SOAP) and representational state transfer (REST) (Halili and Ramadani, 2018; Krimmer et al., 2021; Priisalu and Ottis, 2017). Also, the representation of credentials on the Web should be in a way that is machine-verifiable, private, and cryptographically secure. The semantic interpretation of ID credentials must be unambiguous. That is, verifying credentials and presentations and cryptographically securing them both require predictable, bi-directional, and lossless processes. To be processed in an interoperable manner, any verification of a credential or presentation must be deterministic. The resulting credential or presentation must be semantically and syntactically equivalent to the original construct (Sedlmeir et al., 2021; W3C Recommendation, 2022). Likewise, each verified credential of $U_{ID}$ must be bound to its identity to a given level of assurance. This establishes an unbreakable link between the subject ($U_{ID}$) and the credential to enforce identity disambiguation. Binding is relevant for the verification and validation of the claimed identity at cross-border entry/exit points.

Finally, the security and privacy of the identification data are significant as it forms the integral component of the required trustworthiness for interoperability. There are several approaches for implementing the security and privacy of identification data such as the use of Transport Layer Security (TLS), Security Assertion Markup Language (SAML), authentication keys, etc. (for security), secure computation mechanisms, trusted third party, differential privacy, etc. (for privacy) (Grassi, Garcia and Fenton, 2017; Kaaniche, Laurent and Belguith, 2020). We posit that the use of secure computation mechanisms, data minimisation, and differential privacy in a cross-border context can fulfil the required privacy requirements due to the multifaceted risks associated with the exchange of data between interoperating entities.

## 6.    Conclusion and Future Work

Trustworthy interoperability is a key enabler for identity management and widens the dimensions of e-government in developing countries. This research provided insights into the challenges, opportunities, and requirements of interoperability and highlighted the limitations of current interoperability solutions such as the Estonian X-Road and the eIDAS of the European Union. We found that each existing interoperability solution has its benefits and drawbacks, although both are targeted at a common goal. To achieve seamless communication of identification data, we proposed an approach for scoping requirements by first identifying the key processes to identity management in an interoperability context. Furthermore, we addressed the limitations of the current solutions using an architecture that considers the flexibility of identity verification and validation through an interoperable layer that is based on four interacting layers of data representation, semantics, binding, and security and privacy. For future work, an analysis of country-specific use cases for achieving interoperability in the African identity ecosystem will be performed.